\documentclass[fleqn,10pt]{wlscirep}
\usepackage[utf8]{inputenc}
\usepackage[T1]{fontenc}
\title{Mxenes for CO$\rm_{2}$ reduction and catalytically improved liquid hydrogen storage  vie reverse water gas shift reaction.}

\author[1]{T.E.Ada}
\author[2]{K.N.Nugussa}
\author[3]{Cecil~N.M.~Ouma}

\affil[1, 2]{Department of Physics,\ Addis Ababa University,\ P.O. Box 1176,\      Addis Ababa,\ Ethiopia}  
\affil[3]{HySA Infrastructure Center of Competence, Faculty of Engineering, North-West University, Private Bag X6001, Potchefstroom 2520, South Africa}

\affil[*]{corresponding.moronaphtaly84@gmail.com\ 
               (Cecil~N.M.~Ouma)}


\begin{abstract}
The catalytic reduction of $\mathrm{CO_{2}/CO}$ is an appealing approach for reducing greenhouse gas concentrations while also producing renewable energy. We used two-dimensional transitional metal carbides known as Mxenes as the most promising catalysts for boosted water-gas-shift reaction for conversion of $\mathrm{CO_{2}}$ to chemical fuel and liquid hydrogen. Our findings reveal that the $\mathrm{Ti_{2}C}$ surface collects         $\mathrm{CO_{2}}$ and converts it to reactive carbon mono oxide gas and oxygen termination. Surface catalytic reactions always start with $\mathrm{CO}$ hydrogenation, which is sustained by a continual supply of water at the optimum temperature.  
  $\mathrm{Ti_{2}C}$ surface terminations are in charge of the formation of molecules, free radicals, and alcohols, and the conversion reaction is cycled frequently, producing methanol, methane, water, and hydrogen molecules with each cycle. Furthermore, once water is injected for system hydrogenation, the $\mathrm{Ti_{2}C}$ surface has the ability to hydrogenate itself, because water breaks down into its constituents $\mathrm{O}$ and $\mathrm{OH}$ in the presence of free radicals such as $\mathrm{H_{2}CO}$. Thus, self hydrogenation increases liquid hydrogen generation in addition to the usage of water for hydrogen supply.

\end{abstract}
\begin{document}

\flushbottom
\maketitle
%
%
\thispagestyle{empty}


\section*{Introduction}

Due to their numerous applications, 2D materials have emerged as a cutting-edge study field. The newest developments in the 2D universe include transition metal carbides, carbonitrides, and nitrides (MXenes). MXenes are composed of layers of transition metal carbides or nitrides, $\mathrm{M_{n+1}X_{n}}$ where  $\mathrm{M}$ stands for a transition metal (Sc, Ti, Zr, Hf, V, Nb, Ta, Cr, Mo, and so on), and X is either carbon or nitrogen. Various Mxene characteristics, from metallic to semiconductor, depend on the nature of M, X, and surface termination. High mechanical stability, high electronic conductivity, chemical stability, ion intercalation, and tunable band gaps are just a few of the Mxenes' many desirable properties. These properties have important implications for energy applications like fuel cells, hydrogen storage, and lithium ion batteries. Furthermore, encouraging outcomes have been attained in a variety of fields, including spintronics, wearable electronics, environmental remediation, and biomedicine. 

Mxenes are enticing prospects for catalysis because of the strong reactivity of the resulting surface and the naturally large area of these materials. Mxenes meet several of the characteristics that are frequently needed for catalysts to be stable under reaction conduction and selectivity of reaction product. The exothermic adsorption of MXenes suggests that they could be used as CO$\rm_{2}$ conversion catalysts, and they obviously show promise as materials for CO$\rm_{2}$ capture.

\begin{figure}[htpb!]
        \centering
        \includegraphics[scale=0.25]{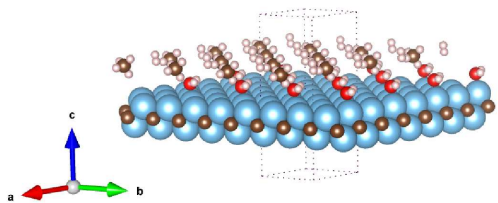}
        \caption{\label{fig1} Schematic of syngas catalytic activity on Mxene, $\mathrm{Ti_{2}C}$ surface, color code: silver color for Titanium, transition metal element, brown for Carbon, red for Oxygen, and white for Hydrogen.}
\end{figure}

The over-reliance on fossil fuels in modern society has negative effects on climate change and ocean acidification. These two phenomena have a strong correlation with CO$\rm_{2}$ levels in the Earth's atmosphere. A currently popular option for minimizing these effects is the employment of solid collectors to reduce the CO$\rm_{2}$ concentration in accordance with the so-called carbon capture and storage (CCS) plan. Since CO$\rm_{2}$ has a high degree of chemical stability and only weakly interacts with most substrates, thus, active substrates for the selective, robust adsorption of CO$\rm_{2}$ are necessary for this. To put it another way, CO$\rm_{2}$ capture is conceptually straightforward, but really putting it into practice on a large enough scale is rather difficult.

Several attempts have been made over the years to address this issue, and Jorchick \emph{et al} proposed a two-in-one reactor with the advantage of performing both hydrogenation and dehydrogenation save catalyst and equipment costs for stationary energy storage applications in remote areas. For example, simply altering the pressure and the dehydrogenation reaction could be switched to the hydrogenation reaction using the same catalyst~\cite{JPDSMBW_2017}.

The most inefficient technique is to use the electrical output from a proton exchange membrane fuel cell~(PEMFC) to dehydrogenate a charged LOHC system. A hydrogen burner, however, can be used to provide the dehydrogenation heat. A portion of the hydrogen produced by a liquid organic hydrogen carrier~(LOHC) is used to heat the dehydrogenation system, while the remainder is used to power the PEMFC. This technique has a maximum efficiency of roughly 34~$\%$, making it preferable to electrical heating~\cite{PPPCWP_2017}.

$\mathrm{CO_{2}}$ reduction by reverse water gas shift~(RWGS), on the other hand, creates a gas mixture composed of $\mathrm{CO}$ and $\mathrm{H_{2}}$, which is commonly referred to as synthesis gas or syngas. The $\mathrm{CO_{2}}$ dissociation is regarded the rate-determining phase in the RWGS reaction, and its dissociative adsorption heat over the metal governs the reaction rate. The RWGS requires a very high temperature for thermodynamic equilibrium, whereas the chemical equilibrium is pressure independent. Furthermore, the chemical kinetics of the reactions are heavily influenced by the catalyst structure, gas composition, and coating processes~\cite{GMDBAH_2021}.

Reverse water gas shift catalysis looks to be the dominant method for converting $\mathrm{CO_{2}}$ to $\mathrm{CO}$, which can subsequently be transformed to liquid fuel of choice via $\mathrm{CO}$ hydrogenation, such as diesel, gasoline, and alcohols. This technique has the advantage of high rates, selectivity, and technological readiness, but it requires renewable hydrogenation via direct photocatalysis or indirect sources such as electricity and electrolysis.

The synthesis of liquid hydrogen fuel from $\mathrm{CO_{2}/CO/H_{2}}$ feeding is a significant method that is heavily reliant on reaction circumstances and the type of catalyst used. The liquid yields, such as liquid hydrogen is far more advantageous in terms of reducing the shortcoming of fossil fuels. $\mathrm{CO_{2}/CO/H_{2}}$ is used in the reaction technique to synthesis hydrogen fuel, which is a reverse process of hydrogen fuel reforming at the same time. When $\mathrm{H_{2}}$ is introduced for material reduction, a reverse water-gas shift chemical looping reaction occurs~\cite{GVPHBVDCMG_2013}.

The reversible hydrogenation of $\mathrm{CO_{2}}$ to create $\mathrm{CO}$ and $\mathrm{H_{2}O}$ is represented by the RWGS reaction (Eq.~\ref{eq1}). $\mathrm{CO_{2}}$ is a relatively nonreactive molecule due to its chemical stability, and hence the reaction to convert it to the more reactive $\mathrm{CO}$ is energy intensive.

\begin{eqnarray}
\label{eq1}
& \mathrm{CO_{2}}+\mathrm{H_{2}} \rightleftharpoons \mathrm{CO} + \mathrm{H_{2}O}, \mathrm{\Delta H^{\circ}_{298~K}=+42.1~~\frac{KJ}{mol}}\\
& \mathrm{CO_{2}}+\mathrm{4H_{2}} \rightleftharpoons \mathrm{CH_{4}} +\mathrm{2H_{2}O}, \mathrm{\Delta H^{\circ}_{298~K}=-165~\frac{KJ}{mol}}
\label{eq2}
\end{eqnarray}

Since the reaction is endothermic, greater temperatures are thermodynamically advantageous. Increased $\mathrm{H_{2}/CO_{2}}$ ratio maximizes $\mathrm{CO_{2}}$ conversion and favors the RWGS reaction~\cite{DYKJ_2016}. As a result, at lower temperatures, the equilibrium will gradually favor the reverse reaction of Eq.~(\ref{eq1}) and methanation, Eq.~(\ref{eq2}) reactions, which are exothermic and the most notable side reactions under these conditions. However, this ratio and temperature are restricted to ensure that the parameters used during the experimental stages are economically viable for industrial applications. Previous research has demonstrated that changing the pressure has no effect on the reaction activity and position of the equilibrium due to the stoichiometry of the reaction~\cite{LJWMMNLR_2016}.

Two temporal steps can be recognized in the RWGS process: reduction of Titanium carbide with $\mathrm{H_{2}}$ followed by re-oxidation with $\mathrm{CO_{2}}$ generates the target product, $\mathrm{CO}$. The streamlined reaction scheme

\begin{eqnarray}
\centering
\text{Reduction}&  \mathrm{Ti_{2}C}+\mathrm{2H_{2}} \rightarrow \mathrm{2Ti} + \mathrm{CH_{4}}\\
\text{Oxidation}&  \mathrm{2Ti}+\mathrm{2CO_{2}}+\mathrm{CH_{4}} \rightarrow \mathrm{Ti_{2}C} + \mathrm{2CO}+\mathrm{2H_{2}O}
\end{eqnarray}

Direct $\mathrm{CO_{2}}$ hydrogenation is more thermodynamically favorable than RWGS and thus more promising for industrialized methanol synthesis; however, $\mathrm{CO_{2}}$ hydrogenation to generate methanol via reverse-water-shift process found 20$\%$ greater methanol yields when $\mathrm{CO_{2}}$ is converted to $\mathrm{CO}$ rather than $\mathrm{CO}$ to methanol using direct $\mathrm{CO_{2}}$ hydrogenation~\cite{JOJKMRALGSUS_1999}.

Catalysts such as Fe, Co, Ni, Cu, or noble metals are used depending on the end product. Because $\mathrm{CO}$ generated from $\mathrm{CO_{2}}$ is more reactive than $\mathrm{CO_{2}}$, it can engage in additional hydrogenation processes. Thus, controlling product selectivity in $\mathrm{CO_{2}}$ hydrogenation to noble metal-based substrates is highly problematic due to multi-faced reaction network, which includes three major reactions: $\mathrm{CO_{2}}$ reduction to $\mathrm{CO}$ in the reverse water gas shift process, CO hydrogenation to methanol, and $\mathrm{CO}$/$\mathrm{CO_{2}}$ hydrogenation to $\mathrm{CH_{4}}$.

\section*{Results and Discussion}
\begin{figure}[htpb!]
        \centering
        \includegraphics[scale=0.25]{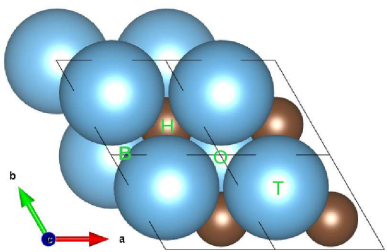}
        \caption{\label{fig2} Top views of $\mathrm{Ti_{2}C(2 \times 2 \times 1)}$ clean surface, and B, O, T $\&$ H denotes possible adsorption sites.}
\end{figure}

\begin{figure*}[htpb!]
        \centering
        \includegraphics[scale=0.55]{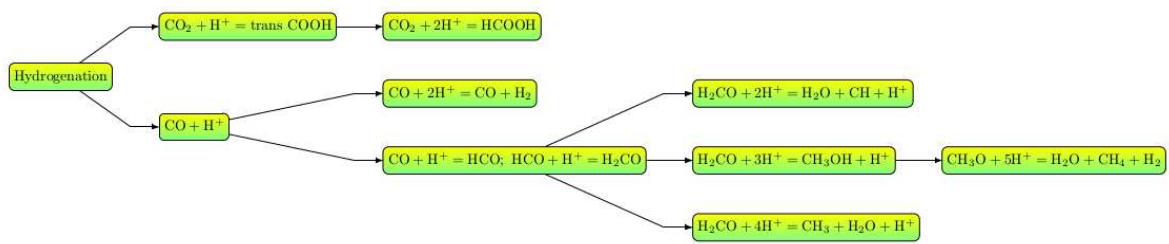}
        \caption{\label{fig3} Flow chart of step by step hydrogenation of $\mathrm{CO_{2}}$ and $\mathrm{CO}$, selectivity of catalyst building stable intermediates or molecules, and  preferred reduction reaction path way for liquid Hydrogen production.}
\end{figure*}

We investigated decomposition or formation of adsorbant or intermediates on the clean $\mathrm{Ti_{2}C(2 \times 2 \times 1)}$ surface of which active site indicted on Fig~.\ref{fig2}, and B:~bridge site, O:~off-site, T:~top site and H:~hallow site. We studied step by step hydrogenation of $\mathrm{CO_{2}}$ or $\mathrm{CO}$, and the reaction schemes are shown on Fig.~\ref{fig3}. Here we discussed Mxenes~\cite{MDACVFIF_2020} importance in enhancing catalysis reaction in relation to adsorption, activation energy, and Bronsted-Erans-Polyani(BEP) relation describes the linear relation which exists between activation energy $\&$ reaction energy~\cite{C8CP01476K}. 

Electrochemical reduction of CO$\rm_{2}$, often surface properties determines the main product of CO$\rm_{2}$ reduction~\cite{BAJWVSSPRJ_2017}, hence $\mathrm{Ti_{2}C}$ surface absorbs CO$\rm_{2}$ at bridge site which is highly reactive, thus, it dissociate to $\mathrm{CO}$ and $\mathrm{O}$, and this turns on reverse-water-gas-shift reaction with hydrogenation of $\mathrm{CO}$, resulting in subsequent chain reactions. 

The reduction of $\mathrm{CO_{2}}$ to $\mathrm{CO}$ on Ti-metal surface experimentally verified in litureature Ref.\cite{BACOIVRJ_2022} which agrees with our work. The $\mathrm{Ti_{2}C}$ bridge site is highly reactive and enough to dissociate or break the binding energy of $\mathrm{CO_{2}}$ into $\mathrm{CO}$ and $\mathrm{O}$ with energy, 4.09~eV, the binding energy (BE) of surface intermediates was determined according to the following equation,
\begin{equation}
\mathrm{BE_{x}=E_{slab+x}-E_{slab}-E_{x}}
\end{equation}
Where $\mathrm{E_{slab+x}}$, $\mathrm{E_{slab}}$ and $\mathrm{E_{x}}$ are the total energies of the adsorbate plus slab system, the clean slab and the gas-phase intermediate, X respectively. According to this equation, a negative value of $\mathrm{BE_{x}}$ signifies a exothermic adsorption. 

Table.~\ref{tab1} demonstrates that hydrogen atom is especially reactive in all sites, which facilitates the start of the water-gas-shift-reaction~\cite{JSMM_2016} process, which uses reactive hydrogen ion quickly after $\mathrm{CO_{2}}$ breakdown into constituents.

The competitive hydrogen evolution reaction occurs at multiple active $\mathrm{Ti_{2}C}$ surface regions, as illustrated in the flow chart.Thus, when reactive $\mathrm{CO}$ and $\mathrm{O}$ react with $\mathrm{H_{2}}$ result in Formic acid, the hydrogen molecule may require sun energy to ionize itself for active reaction, or hydrogenation via the water gas shift reaction is another viable approach for the production of Formic acid, which is a natural fuel. Similarly, sequential $\mathrm{CO}$ hydrogenation results in the formation of methanol, methane, water, and the hydrogen molecule. However, because the oxidation reaction rate drops at low temperatures, the reaction should be controlled by the supply of water and the temperature.

The $\mathrm{Ti_{2}C}$ surface terminations, such as $\mathrm{O}$ and $\mathrm{OH}$, are critical in the formation of reactive intermediates as well as the inhomogeneous charge arrangement on the Mxenes layers, impacting the surface's shape and microstructure. As a result, $\mathrm{CO_{2}}$  is dissociated at the surface  resulting in start of
 conversion reaction, and catalytic cycle must be maintained.  Hydrogen and oxygen have to be produced in a molecular ratio of two, or else the system will become unstable and the catalyst will be consumed during the reaction~\cite{MPOGRWPBD_2019}. Direct hydrogenation from free hydrogen molecules or water at moderate temperatures is thus required for the splitting constituent atoms, culminating in the sequential synthesis of methane, methanol, hydrogen gas, and water.

\begin{table*}[!htbp]
\setlength{\tabcolsep}{2.4mm}
\renewcommand{\arraystretch}{1.5}
\centering
\small
\caption{Calculated  adsorption energy\ 
         of various intermediates and adsorbant on bridge, hallow, off-site and on top site of \ Ti$\rm_{2}$C  using\ 
         van der Waals exchange functional: optPBE$-$vdw.}
\centering
\begin{tabular}{c c c c c c}
\hline
 &  & & Adsorption energy sites &  & \\
Adsorbant  &  Bridge[eV]& Hallow[eV] & Off-site[eV] & On-top[eV] \\  \hline
 
CO &  -2.991 &-1.915& -1.702  & -1.456 \\
              
            CO$_{2}$ & $\mathrm{CO+O}$,  & $\mathrm{CO_{2}}$,  & $\mathrm{CO_{2}}$, & $\mathrm{CO_{2}}$,\\
            & -4.09&-0.482 &-0.293&-0.331 \\
          
 H$_{2}$&  -0.021& -0.022 & -0.022 &-0.021 \\
 H$_{2}$O&  -0.888& -0.219 & -0.191 &-0.163 \\           
 CHO&  -5.074& -5.073 & -5.065 &-5.07 \\
 $\mathrm{cis-COOH}$&  $\mathrm{cis-COOH}$, & $\mathrm{cis-COOH}$, & $\mathrm{CO+OH}$, &$\mathrm{CO+OH}$, \\
 &  -4.386& -4.940 & -7.016 &-7.014 \\
 H&  -4.530& -4.530 & -4.274 &-3.061 \\
 O&  -8.673& -8.675 & -8.575 &-6.339 \\
 OH& $\mathrm{O+H}$,  & $\mathrm{OH}$, & $\mathrm{OH}$, &$\mathrm{OH}$, \\
 &  -6.886& -3.776 & -4.047 &-0.989 \\ 
Intermediates &   &  &  & \\
\hline
 CH& $\mathrm{C+H}$, & $\mathrm{CH}$, & $\mathrm{CH}$, & $\mathrm{CH}$,\\
 &  -6.39& -4.74 & -4.60 &-1.42 \\
$\mathrm{CH_{3}}$&  -3.24& -3.18 & -2.77 &-1.73 \\
$\mathrm{CH_{4}}$&  -0.16& -0.16 & -0.16 &-0.17 \\
$\mathrm{CH_{3}O}$&  -5.04& -5.10 & -5.04 &-5.11 \\
$\mathrm{CH_{3}OH}$& $\mathrm{CH_{3}OH}$, & $\mathrm{CH_{3}O+H}$, & $\mathrm{CH_{3}OH}$, &$\mathrm{CH_{3}OH}$, \\
&  -0.87& -4.12 & -0.75 &-0.75 \\
$\mathrm{H_{2}COH}$& $\mathrm{CH_{2}+OH}$,  & $\mathrm{H_{2}COH}$ & $\mathrm{CH_{2}+OH}$, &$\mathrm{CH_{2}+OH}$, \\
&  -5.71& -3.20 & -5.95 &-3.20 \\
 $\mathrm{HCOOH}$&  $\mathrm{HCOOH}$,  &  $\mathrm{HCOOH}$, &  $\mathrm{CHO+OH}$, &  $\mathrm{HCOOH}$,\\
 &  -2.34& -2.67 & -4.59 &-2.48 \\
 \hline
\end{tabular}
\label{tab1}
\end{table*} 

In our scenario, continuous hydrogenation is necessary since the reaction cycle would not be sustained; this is mostly owing to hydrogen supply from water split, where continual watering in the presence of free radicals such as $\mathrm{H_{2}CO}$ leads to water splitting into $\mathrm{H}$  and $\mathrm{OH}$ as one can note from Table.~\ref{tab2}. This stage is critical because constant supply of $\mathrm{OH}$ termination comes from water, and $\mathrm{OH}$ is responsible for the creation of intermediates as well as $\mathrm{O}$ termination from $\mathrm{CO_{2}}$ dissociation, both of which are crucial in the stabilization of the $\mathrm{Ti_{2}C}$ surface. As a result, $\mathrm{H}$  and $\mathrm{OH}$ play important roles in cycling the entire water-gas-shift reaction process.
 
\begin{table*}[htbp!]
\setlength{\tabcolsep}{2.3mm}
\renewcommand{\arraystretch}{1.5}
\centering
\small
\caption{Calculated  binding energy\ 
         of anion, cation, and radical interactions on bridge, hallow, off-site and on top site of \ Ti$\rm_{2}$C  using\ 
         van der Waals exchange functional: optPBE$-$vdw.}
\centering
\begin{tabular}{c c c c c c}
\hline
 &  & & Adsorption energy sites &  & \\
Anion-cation interaction  &  Bridge[eV]& Hallow[eV] & Off-site[eV] & On-top[eV] \\  \hline 
 $\mathrm{H+H \rightarrow}$& $\mathrm{H_{2}}$,  & $\mathrm{H_{2}},$ & $\mathrm{H_{2}}$, &$\mathrm{H_{2}}$, \\
 &  -0.015& -0.017 & -0.016 &-0.14 \\
    $\mathrm{C+H \rightarrow}$&  $\mathrm{CH}$,& $\mathrm{CH}$, & $\mathrm{CH}$, &$\mathrm{CH}$, \\
    &  -7.00& -6.96 & -7.00 &-7.00 \\
 $\mathrm{O+H \rightarrow}$&  $\mathrm{OH}$, & $\mathrm{OH}$, & $\mathrm{OH},$ &$\mathrm{OH}$, \\
 &  -5.70& -5.77 & -5.70 &-5.70 \\
 $\mathrm{H+OH \rightarrow}$& $\mathrm{H_{2}O}$,  & $\mathrm{H_{2}O}$, &$\mathrm{H_{2}O}$,  &$\mathrm{H_{2}O}$, \\
 &  -1.03& -0.83 & -0.85 &-0.84 \\
 $\mathrm{OH+OH  \rightarrow }$&  $\mathrm{OH+H+O}$,& $\mathrm{OH+H+O}$ ,& $\mathrm{OH+OH}$, &$\mathrm{OH+OH}$, \\
  &  -9.96 & -9.85 & -8.85 &-8.84 \\
    $\mathrm{H+OH+O \rightarrow}$ &  $\mathrm{H_{2}O+O}$, & $\mathrm{OH+O+H}$, & $\mathrm{H_{2}O+O}$, &$\mathrm{H+O+OH}$, \\
  &  -9.33& -7.81 & -10.57 &-9.71 \\
   $\mathrm{H+OH+O+CO \rightarrow}$& $\mathrm{OH+trans-COOH}$, & $\mathrm{OH+trans-COOH}$, & $\mathrm{HCO+OH+O}$, &$\mathrm{OH+trans-COOH}$, \\
    &  -9.33& -7.81 & -10.57 &-9.71 \\
\hline
    Water splitting &   &  &  & \\

 $\mathrm{H_{2}O+H_{2}CO \rightarrow}$& $\mathrm{H+OH+H_{2}CO}$, & $\mathrm{H+OH+H_{2}CO}$, & $\mathrm{H+OH+H_{2}CO}$, & $\mathrm{H+OH+H_{2}CO}$,\\
 &  -3.69& -4.02 & -4.17 &-6.84 \\
     Free hydrogen molecule &   &  &  & \\

 $\mathrm{H_{2}+H_{2}CO \rightarrow}$& $\mathrm{CH_{3}OH}$, & $\mathrm{CH_{3}OH}$, & $\mathrm{CH_{3}OH}$, & $\mathrm{CH_{3}OH}$,\\
 &  -0.34& -0.08 & -0.37 &-0.28 \\
  
 \hline
\end{tabular}
\label{tab2}
\end{table*}

\section*{Conclusion}
In conclusion, we calculated the multi-active single layered $\mathrm{Ti_{2}C}$ surface using basic principles. We discover that the bridge site is the only location where $\mathrm{CO_{2}}$ dissociates into $\mathrm{CO}$ gas and $\mathrm{O}$ termination, and that the gate for the reverse-water-gas-shift process begins with hydrogen ignition and continues with water supply for hydrogenation and $\mathrm{OH}$ termination. Surface terminations play a critical part in the overall cycling of the reaction route and the creation of liquid hydrogen plus methanol fuel.   Our research reveals that $\mathrm{Ti_{2}C}$ surfaces are capable of self-hydrogenation due to free radicals such as $\mathrm{H_{2}CO}$ generated at the onset of hydrogenation. Hydrogen molecules may be formed during the conversion process from hydrogen ions, but these molecules are promptly consumed in the creation of methanol.

\section*{Methods}
All calculations were performed using the GPAW code. The $\mathrm{Ti_{2}C}$ surface was modeled with a one-layered slab using ($\mathrm{2\times 2\times 1}$) surface unit cell. Plane augmented wave method was applied to describe ionic cores and the Kohn-sham valence electron states were expanded in a basis of plane waves with kinetic energy below 400~eV. The surface Brillouin zone was sampled at $\mathrm{2\times 2\times 1}$ k-points. We implemented optPBE$-$vdw exchange correlation which has much better chemical accuracy than vdW-DF for the S22 benchmark set of weakly interacting dimers and for water clusters with improved performance for the adsorption of water~\cite{JKDBAM_2010}.

\bibliography{sample}
\section*{Acknowledgements}
Financial support from the North-West University~(Potchefstroom Campus),  the DST HySA Infrastructure Centre Competence~(KPs program), and Addis Ababa university are gratefully acknowledge, the Centre for High Performance Computing~(CHPC) in Cape Town~(South Africa) for computational resources used in this study.

\end{document}